\documentclass[reprint,amsmath,amssymb,aps,showkeys,floatfix,prapplied]{revtex4-2}
\usepackage{graphicx}

\begin{document}
\title{Flexural wave modulation and mitigation in airfoils using acoustic black holes}

\author{Kaushik Sampath}\email[]{kaushik.sampath@nrl.navy.mil}
\author{Caleb F Sieck}\author{Matthew D Guild}\author{Alec K Ikei}
\affiliation{U.S. Naval Research Laboratory, Code 7165, Washington DC 20375, USA}
\author{Charles A Rohde}
\affiliation{U.S. Naval Research Laboratory, Code 6364, Washington DC 20375, USA}
\date{\today}
\begin{abstract}
This study introduces a framework for the design and implementation of acoustic black holes (ABHs) in airfoils. A generalized multi-parameter damped-ABH generation function is mapped onto NACA series airfoils. Representative geometries and a uniformly distributed baseline, all with the same mass of structure and damping are fabricated using multi-material PolyJet 3D printing. Laser Doppler vibrometer measurements along the airfoil chord in response to a broadband 0.1 - 12 kHz excitation show a decrease in trailing edge vibrations by as much as 10 dB, a broadband 5 dB reduction across the entire chord as well as substantial spatial and temporal modulation of flexural waves by ABH-embedded foils. Finite element analysis (FEA) models are developed and validated based on the measured data. Furthermore, a parametric FEA study is performed on a set of comparable designs to elucidate the scope of modulation achievable. These findings are applicable to trailing-edge noise reduction, flow control, structural enhancement and energy harvesting for airfoils.
\end{abstract}

\maketitle

\section{\label{intro} Introduction}

Fluid-loaded structures such as turbomachine blades and aircraft wings are often designed slender due to constraints on weight, making them susceptible to vibrational excitation by flow~\cite{rogerBroadbandSelfNoise2004}. This leads to an increased wear of structures, affecting longevity and performance. Decades of past and ongoing research has been aimed at finding better ways to mitigate such undesirable consequences~\cite{brooksAirfoilSelfnoisePrediction1989}. Substantial efforts have also gone into the redistribution and harvesting of flow-induced vibrations for controlling turbulent flow and overall energy efficiency. Trailing edge noise, which is in fact, a subset of the above, still remains an active topic of research due to its relevance to airframes, propellers and rotors~\cite{brooksTrailingEdgeNoise1981,howeReviewTheoryTrailing1978,gruberAirfoilNoiseReduction2012}.

New approaches of structural geometry modifications, such as, applying the so-called acoustic black hole (ABH) effect have become increasingly popular. Mironov~\cite{mironovPropagationFlexuralWave1988} theorized that flexural waves can be ‘trapped’ in a beam with an ideal power law-shaped tapering end. This is because the group velocity goes to zero as it scales with square root of the edge thickness. In practice, this is leveraged by adding viscoelastic damping wherever the taper truncates~\cite{krylovExperimentalInvestigationAcoustic2007,oboyDampingFlexuralVibrations2010,krylovAcousticBlackHoles2014,feurtadoExperimentalInvestigationAcoustic2016,krylovAcousticBlackHoles2004a, climenteOmnidirectionalBroadbandInsulating2013}. A detailed review of ABH theory and applications has recently been carried out by Pelat et al~\cite{pelatAcousticBlackHole2020}. Evidently, despite their popularity, applications of ABHs have been largely restricted to beams, plates and more recently, cylindrical shells~\cite{dengReductionBlochFloquetBending2020}. 

As far as aerodynamic applications are concerned, 1D power-law tapers have been successfully incorporated into the trailing edge of turbo-fan blades~\cite{bowyerDampingFlexuralVibrations2014}. The study found that the measured acceleration, especially around resonances, was substantially reduced in airfoils with power law tapers (ABHs) when compared to those without. As this study notes, fabricating a trailing edge with a power law taper is not trivial, but still achievable in an otherwise seamless manner. However, for other use cases, where flow-structure interactions may need to be modulated in sections of the chord besides the trailing edge, structural modifications would need to be concealed internally without affecting the aerodynamic external shape of the airfoils. 

In fact, only recently have numerical studies even looked at embedding ABHs in higher dimensional closed geometries such as cylindrical shells and beams~\cite{dengTransmissionLossPlates2021,dengSemianalyticalMethodCharacterizing2020,dengReductionBlochFloquetBending2020}. Deng et al.~\cite{dengReductionBlochFloquetBending2020} compare the performance of a uniformly damped cylindrical steel shell with that of a ten-element ABH-embedded shell with the same damping layer thickness. They also evaluate the effect of the truncation thickness and found that even for their thickest (least favorable) truncation case there is a 10 dB reduction in transmissibility of flexural vibrations when compared to the uniformly damped case. It is important to reiterate that unlike the previously mentioned turbo-fan blade edges~\cite{bowyerDampingFlexuralVibrations2014}, cylindrical shells and beams do not have an obvious site to embed ABH tapers and there is a large impact on the overall rigidity of the shells. A work-around proposed by  Deng et al.~\cite{dengReductionBlochFloquetBending2020} is the addition of specially directed `stiffeners' that effectively support the structure without compromising the ABH effect. These numerical studies, to the best of the authors' knowledge, remain the only examples of embedded (or closed-geometry) implementations of ABHs in higher-dimensions. It still remains to be seen whether ABH-embedded designs can be fabricated and demonstrated as such.  

Recent advances in additive manufacturing, such as multi-material PolyJet printing enable rapid fabrication of complex designs with hard and soft materials in a single build. Subsequently, power law tapers, including functionally graded ABHs, have been 3D-printed recently~\cite{huangLowReflectionEffect2019} for beams. Comparisons of the measured reflection coefficients between a traditional (or single material) ABH beam and those spatially distributed with softer and higher loss materials towards the tapering end show an order-of-magnitude reduction in the latter.

Motivated by the above-mentioned works, the present study aims to provide a framework for the design and implementation of ABH-embedded airfoils for the modulation and mitigation of flexural waves. The organization of the remaining sections in this paper is as follows. In Section~\ref{methods}, the methods used in this paper are presented, where, a wide range of ABH parameters are mapped inside an airfoil profile (Sections~\ref{abhgenfun} and ~\ref{abhfoilgeom}) followed by the materials and fabrication of representative geometries with the same total mass of structural and damping material in Section~\ref{matfab}. A description of the experimental setup where Laser Doppler vibrometry (LDV) is used to characterize chord-wise vibrations when the airfoils are subjected to a leading-edge point excitation in the 0.1-12 kHz range is provided in Section~\ref{exptsetup}. Subsequently, FEA-based simulation and modeling is introduced in Section~\ref{modeling}. The results of this work are presented and discussed in Section~\ref{results}, beginning with an examination of the wavenumber-frequency characteristics in Section~\ref{kw_analysis}, followed by chord-frequency characteristics in Section~\ref{cf_analysis}, a detailed discussion of the FEA results in Section~\ref{fea_results} and the ABH airfoil parametric study in Section~\ref{parametric_study}. The conclusions of this work are then summarized in Section~\ref{conclusions}.

\section{\label{methods}Methods}
\subsection{\label{abhgenfun}ABH generating function}
\begin{figure}
\begin{center}\includegraphics{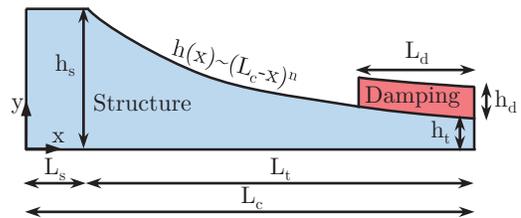}\end{center}
\caption{Schematic defining input parameters for the ABH generating function}
\label{figure_nomenclature}
\end{figure}

\begin{figure}
\begin{center}\includegraphics{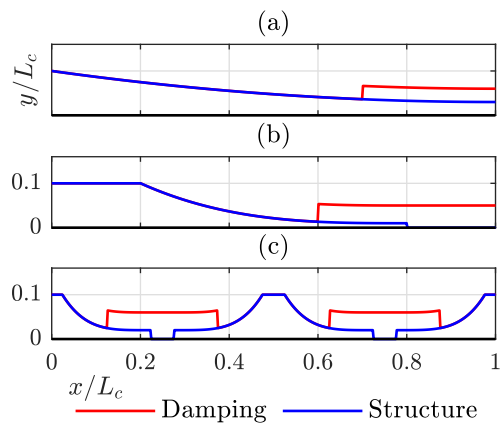}\end{center}
\caption{Sample geometries from the ABH generating function for (a) $L_\text s$=0, $L_{\text t}$=1, $h_{\text t}/h_\text s$=0.3, $n$=2, $N$=$1/2$, $L_{\text d}$=0.3, $h_{\text d}/h_\text s$=0.3, (b) $L_{\text s}$=0.2, $L_{\text t}$=0.6, $h_{\text t}/h_\text s$=0.1, $n$=3, $N$=$1/2$, $L_{\text d}$=0.4, $h_{\text d}/h_\text s$ =0.4 and (c) $L_{\text s}$=0.1, $L_{\text t}$=0.8, $h_{\text t}/h_\text s$=0.2, $n$=4, $N$=2, $L_{\text d}$=0.5, $h_{\text d}/h_\text s$=0.4.}
\label{figure_genfun}
\end{figure}

An ABH generating function is formulated with several parameter inputs (Figure~\ref{figure_nomenclature}). These have been chosen after compiling recent works on optimization of the ABH shape or ABH-embedded shells and beams~\cite{mccormickDesignOptimizationPerformance2020,dengSemianalyticalMethodCharacterizing2020,dengTransmissionLossPlates2021}. Horizontal ($x$) and vertical ($y$) axes are defined along the length and thickness of the sample respectively. The thickness, $h$, of the taper changes from its starting maximum value, $h_{\text s}$, to its minimum truncated value, $h_{\text t}$, following a taper power, $n$. The length over which $h$ = $h_{\text s}$ (constant) is denoted $L_{\text s}$, and the taper length over which $h_{\text s}\geq h\geq h_{\text t}$ is denoted $L_{\text t}$. The total length is set to a constant $L_c$, prescribing $L_{\text s}$+$L_{\text t}\leq L_c$, where the equality and inequality represent `continuous' and `truncated' ABHs, respectively. The damping layer is distributed in the direction of increasing thickness from the end. The total length of the damping is denoted $L_{\text d}$ and its height from the taper is denoted $h_{\text d}$. When the ABH is truncated, there exists a damping layer with a total height of $h_{\text t} + h_{\text d}$ in the truncated region to ensure continuity of the exterior. The number of ABHs, $N$, can also be varied by adopting a unit cell approach on the entire length. A single taper is designated by $N$ = $1/2$, and whole numbers represent even number of tapers. The range of values adopted for the different parameters is listed in Table~\ref{table_pars} with representative cases illustrated in Figure~\ref{figure_genfun}. Length and thickness are normalized by $L_c$.

\begin{table}
\caption{Range of input parameters for the ABH generating function}
\label{table_pars}
\centering\begin{ruledtabular}\begin{tabular}{c|ccccccc}
 		&$L_{\text s}$		&$L_{\text t}$		&$h_{t}/h_{\text s}$		&$n$		&$L_{\text d}$		&$h_{\text d}/h_{\text s}$		&$N$\\
\hline
Min			&0.0			&0.8			&0.1				&2		&0.1			&0.0				&$1/2$\\
Step			&0.1			&0.1			&0.1				&1		&0.1			&0.1				&$1/2$\\
Max			&0.2			&1-$L_{s}$	&0.3				&5		&0.5			&0.5				&5.0
\end{tabular}\end{ruledtabular}\end{table}

\subsection{\label{abhfoilgeom}ABH-embedded airfoil geometry}
\begin{figure}
\begin{center}\includegraphics[width=\linewidth]{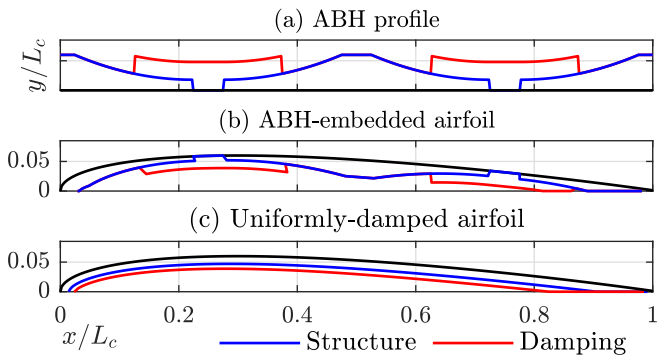}\end{center}
\caption{(a) ABH generating function geometry, (b) ABH-embedded NACA0012 profile and (c) corresponding uniformly distributed baseline profile for $L_\text s$=0.1, $L_{\text t}$=0.8, $h_{\text t}/h_\text s$=0.3, $n$=2, $N$=$2$, $L_{\text d}$=0.5, $h_{\text d}/h_\text s$=0.5}
\label{figure_abhfoil}
\end{figure}

Due to its prevalence in the aerospace community, the National Advisory Committee for Aeronautics (NACA) system of 4-digit airfoil geometries is used. Specifically, the symmetric NACA0012 foil is chosen in this study due to its ubiquity, although the framework can be extended to any shape. Following Ladson et al.~\cite{ladsonComputerProgramObtain1996}, Equation~\eqref{naca00tt} is used to generate the ordinates of a NACA00tt foil, where $tt$ is the chord to thickness ratio.
\begin{equation}\frac{y (tt, x)}{0.05tt} = 0.30\sqrt{x} - 0.13x - 0.35x^2 + 0.28x^3 - 0.10x^4\label{naca00tt}\end{equation}
The ABH profile obtained from the generating function is applied as a normal offset curve~\cite{duchateaujossOffsetCurve2020} on the interior of the foil starting from the leading edge (LE) to the trailing edge (TE) of the foil. The foil shape adds constraints on the ABH-embedded foil in some cases. For instance, the foil chord length, $L_c$ prescribes the maximum thickness of the foil, which is 0.12$L_c$ for the NACA0012, thereby constraining $h_d$. ABH profiles that fall below the $x$ axis after the offset curve calculation are bumped back to the axis. The masses (i.e. areas under the curves) of the structural and damping materials are then calculated and used to generate a baseline geometry where the structural and damping profiles are at a constant (uniform) offset from that of the foil external shape as illustrated in Figure~\ref{figure_abhfoil}.

Based on the parameter space (Table~\ref{table_pars}), a look-up-table (LUT) of ABH-embedded foil designs is created. The target mass of structural and damping material (arbitrarily chosen) is used to shortlist designs that have the same mass (or within a small percentage). Note that the LUT shortlist sizes can be made arbitrarily large by refining parameter increments (Table~\ref{table_pars}).

\subsection{\label{matfab}Materials and fabrication}
The Stratasys J750 PolyJet printer is used for fabrication. It is capable of building prints with multiple hard plastics, soft rubber-like materials as well as a large number of composite materials by combining hard and soft materials. Recently, Huang et al.~\cite{huangLowReflectionEffect2019} used a predecessor of this printer to successfully fabricate and demonstrate the ABH effect in beams. Following their selection, the hard plastic - VeroGray (RGD850) is chosen as the structural material and the soft rubber-like TangoPlus (FLX930) is chosen as the damping material for fabricating ABH-embedded foils. These are printed in a single build using the `high mix' build mode characterized by a build layer resolution of 27 $\mu$m. 

Although these materials have widely been used for various applications, including the fabrication of ABH beams, their stiffness and loss properties are not well established close to the frequency range of the current interest. Huang et al.~\cite{huangLowReflectionEffect2019} perform tests around the 20 kHz range, very close to the present range (0.1-12 kHz), however, due to the lack of available data and to focus primarily on the influence of a Young’s modulus gradient on the wave attenuation, they assume a constant loss factor of 0.1 for all their materials. To better characterize the complex moduli of the 3D printed materials for this study, an ad-hoc non-destructive testing (NDT) technique using commercial grade compressional and shear wave transducers was developed~\cite{guildBroadbandUltrasonicCharacterization2021}. The resulting complex moduli, $E$ the Poisson's ratio, $\nu$, as well as the density, $\rho$ are presented in Table~\ref{table_materials}.

\begin{table}
\caption{Mechanical properties of 3D printed materials}\label{table_materials}
\centering\begin{ruledtabular}\begin{tabular}{c|ccc}
Material	&$E$	[GPa]		&$\rho$[g/cm$^3$]		&$\nu$\\
\hline
VeroGray 	&2.5			&1.16		&0.35\\
TangoPlus &0.65 + 0.4i	&1.18		&0.47
\end{tabular}\end{ruledtabular}\end{table}

To demonstrate the potential of ABH-embedded foils in modulating and mitigating chordwise vibrations, four geometries are fabricated in the same multi-material print job, with varying ABH-generating functions as shown in Table~\ref{table_fab}. The selection was made to ensure a wide range in measured performance while specifically evaluating the effect of truncation ($L_{\text s}$+$L_{\text t}\leq L_{\text c}$) and number of ABH elements ($N$). All cases have $h_{\text t}/h_{\text s}$ = 0.2 and $n$ = 2, i.e. values that have been considered in other studies as well~\cite{dengReductionBlochFloquetBending2020,mccormickDesignOptimizationPerformance2020}. It must however be noted that this selection is not otherwise optimized. It serves as a proof of concept that also validates subsequent FEA models (Sections~\ref{modeling} and ~\ref{fea_results}) and measured material properties (Table~\ref{table_materials}), based on which a parametric study is discussed at the end in Section~\ref{parametric_study}. 
\begin{table}
\caption{3D printed ABH-foil parameters}
\label{table_fab}
\centering\begin{ruledtabular}\begin{tabular}{c|ccccccc}
\# 		&$L_{\text s}$		&$L_{\text t}$		&$h_{\text t}/h_{\text s}$		&$n$		&$L_{\text d}$		&$h_{\text d}/h_{\text s}$		&$N$\\
\hline
1			&0.10		&0.90		&0.20			&2		&0.50		&0.50			&1\\
2			&0.08		&0.87		&0.20			&2		&0.50		&0.50			&3\\
3			&0.12		&0.87		&0.20			&2		&0.50		&0.50			&1\\
4			&0.03		&0.97		&0.20			&2		&0.50		&0.54			&3
\end{tabular}\end{ruledtabular}\end{table}

All fabricated samples have the same total mass of structural and damping material. A fifth baseline sample is also fabricated with the same masses uniformly distributed following the outer shape of the airfoil, as if one were to make a hollow damped version without any ABH-inspired tapers. This is a crucial aspect that has often been overlooked in prior studies. For instance, most studies compare the vibrational response of a beam or plate to that with the so-called ABH version where substantial mass has been removed in machining out the taper~\cite{pelatAcousticBlackHole2020}. While Deng et al.~\cite{dengReductionBlochFloquetBending2020} partly address this by applying the same damping layer thickness for their baseline and ABH-embedded shells, as also noted by them, the difference in structural masses evokes a different response, complicating comparative performance assessment. 

\begin{figure}
\begin{center}\includegraphics[width=2.75in]{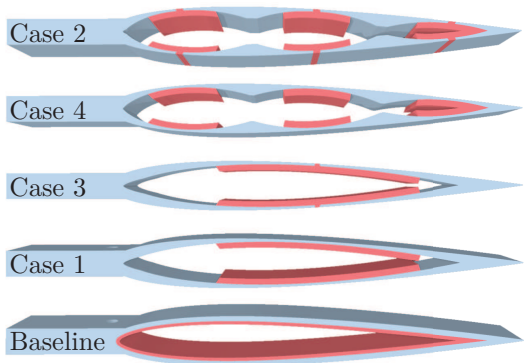}\end{center}
\caption{Fabricated ABH-embedded foil designs. VeroGray and TangoPlus are shaded in blue and pink respectively.}
\label{figure_samples}
\end{figure}

The airfoils have a chord,  $L_c$ = 203.2 mm, resulting in a maximum thickness of 0.12$L_c$ = 24.4 mm. To allow for clamping, a $L_c$/4 = 50.8 mm long section of constant thickness, $L_c/16$ = 12.7 mm is added upstream of the leading edge. Foils are extruded to a depth of $L_c/8$ = 25.4 mm. Images of the fabricated designs are shown in Figure~\ref{figure_samples}. The five fabricated samples weighed 86.2 g on average with a standard deviation of 0.2 g (0.2\%).
\subsection{\label{exptsetup}Experimental setup}
\begin{figure}
\begin{center}\includegraphics{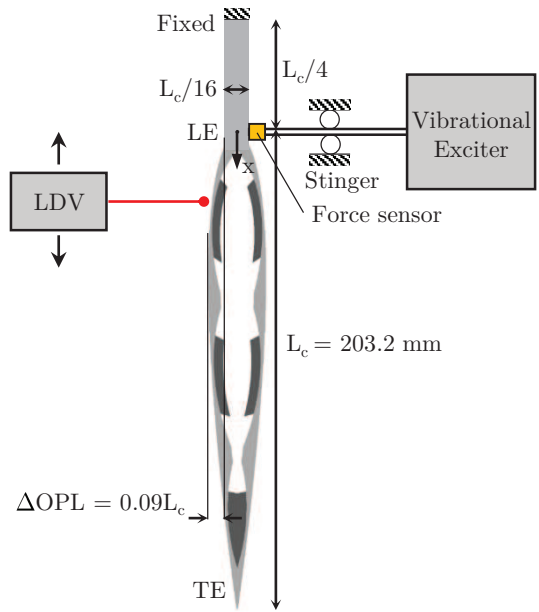}\end{center}
\caption{\label{figure_schematic_bc}Schematic of the experimental setup}
\end{figure}

The ABH-embedded foils are fixed as shown in Figure~\ref{figure_schematic_bc}. 
Vibrational excitation (B\&K Type 4809) over a frequency range of 0.1 - 12 kHz is provided through a stinger to the foil around its leading edge. A piezoelectric force sensor (Model 208A11, 112 mV/N, PCB Piezotronics) is connected between the stinger and foil, allowing an accurate measurement of the force input by the exciter. The vibrational response, i.e. velocity along the chord is measured using a single-point LDV (Polytec CLV-2534). The LDV is mounted on a motorized translation stage (Velmex XSlide) and acquires data over vertical increments of 0.1 mm. Velocity and signal strength from the LDV, as well as force are sampled at a rate of 200 kHz using a National Instruments compactRIO 9035 chassis equipped with analog (NI-9223) and digital (NI-9402) input modules. More details of this setup, including its remote operation have been described by Ikei~\cite{ikeiRemoteOperationSinglePoint2021}.

\begin{figure}
\begin{center}\includegraphics[width=\linewidth]{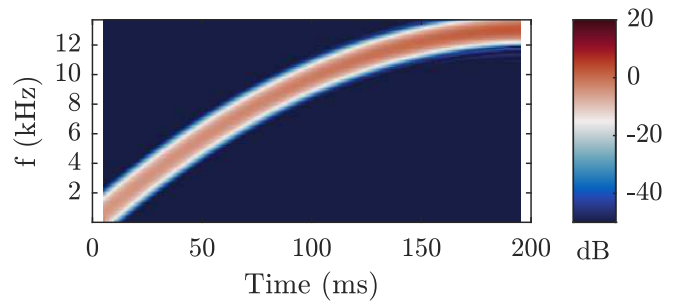}\end{center}
\caption{Spectrogram of the excitation input waveform}
\label{figure_chirp}
\end{figure}

\begin{figure}
\begin{center}\includegraphics{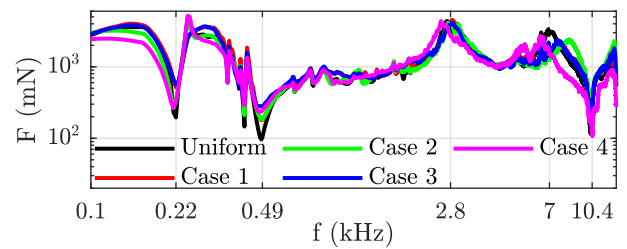}\end{center}
\caption{Frequency spectrum of the measured force input.}
\label{figure_force}
\end{figure}

A 200 ms long chirp excitation signal spanning a frequency range 0.1-12 kHz is sent using a function generator (Agilent 33500B) to a power amplifier (B\&K Type 2718) and serves as the input to the exciter. At each sampling location, LDV data from 32 time sequences is averaged. Given the large input excitation frequency range, a quadratic convex chirp is amplitude-weighted towards higher frequencies as shown in Figure~\ref{figure_chirp}.
The time-averaged force response at the point of excitation is shown in Figure~\ref{figure_force} for all five samples. 
The horizontal and vertical axes are plotted in logarithmic scale denoting frequency, $f$ in kHz and force, $F$ in mN respectively. For convenience, frequency grid lines are chosen near local extrema. Evidently, the same modal signatures, and profiles are found across the entire range for all the cases. The frequency spectra of the LDV data (not shown) calculated at excitation ($x/L_c$=0) are also consistent with the trends in Figure~\ref{figure_force} for all cases.  The forces are as high as 2-3 N around 100-200, 250-400 and 2830 Hz. Minima around  220, 485, and 10445 Hz have forces in the 0.1-0.2 N range, that are presumably anti-resonances associated with structural modes common to all the samples. Consequently, the measured force and velocity (from LDV) at these frequencies are extremely small, comparable to the noise level. However, on either sides of these minima, the force recovers to a value above 1 N. Therefore, it can be concluded that the amplitude weighting is for the most part, effective at preventing any significant decay in the forcing with an increase in frequency, allowing maximum utilization of the LDV's dynamic range. For subsequent analysis, frequency spectra are normalized by the value at excitation, allowing direct inter-frequency and inter-sample comparisons to be drawn. 

As evident from Figure~\ref{figure_schematic_bc}, the foil surface is curved in the direction of the LDV laser beam leading to variations in the optical path length ($\Delta$OPL) of 0.09$L_c$ = 18 mm. The LDV is positioned such that its nearest visibility maximum is centered within $\Delta$OPL leading to a 9\% variation around the peaks that are 204 mm apart. Hence, this is not expected to have a substantial impact on the measurements. Furthermore, the LDV signal strength is also acquired for every measurement, based on which outliers in the data are identified and flagged for subsequent processing steps. The raw LDV data is analyzed in the frequency-wavenumber ($f$-$k$) space to identify bounds for most of the energy content. Subsequently, a Tukey window is applied on the data to remove noise associated with high wavenumbers. A threshold condition of $|k|/2\pi<20/L_c$ is used for all the cases. For convenience, Table~\ref{table_spectral} enlists relevant spectral parameters for space and time.
\begin{table}
\caption{Relevant space and time spectrum parameters}
\label{table_spectral}
\centering\begin{ruledtabular}\begin{tabular}{c|cccc}
	&Samples	&Resolution	&  Max range	&Range of interest\\
\hline
$t$	&40,000		&5 $\mu$s		&0 - 100 kHz		&0.1 - 12 kHz\\
$x$	&3,001		&0.1 mm		&0 - 5 km$^{-1}$	&0 - 618 m$^{-1}$\\
\end{tabular}\end{ruledtabular}\end{table}
\subsection{\label{modeling}Modeling}
Three-dimensional finite element analysis using COMSOL Multiphysics is performed on the five measured cases. In practice, most airfoil wings are thin and hollow, lending themselves to a 2D plate approximation, making FEA substantially less computationally intensive. The present samples also are thin and plate-like in the tapering part of the geometry ($x/L_c\geq$ 0.03 - 0.12, refer Table~\ref{table_fab}). However, clamping and excitation requirements force a beam-like structure upstream ($x/L_c\leq$ 0) making the present structures complex and requiring a 3D model for accurate results. The CAD model used for fabricating the foils is directly imported into COMSOL for analysis. A fixed boundary condition is applied at the upstream edge (Figure~\ref{figure_schematic_bc}). A force of 1 N, based on the force measurements (Figure~\ref{figure_force}) is distributed where the force sensor mounts to the foil. The computational domain is halved by leveraging symmetry in the direction of extrusion to reduce solver time. Material properties as per Table~\ref{table_materials} are applied to the model. A frequency domain simulation spanning the 0.1 - 12 kHz range (Table~\ref{table_spectral}) with a resolution of 100 frequencies per decade is executed.

\begin{figure}
\begin{center}\includegraphics[width=\linewidth]{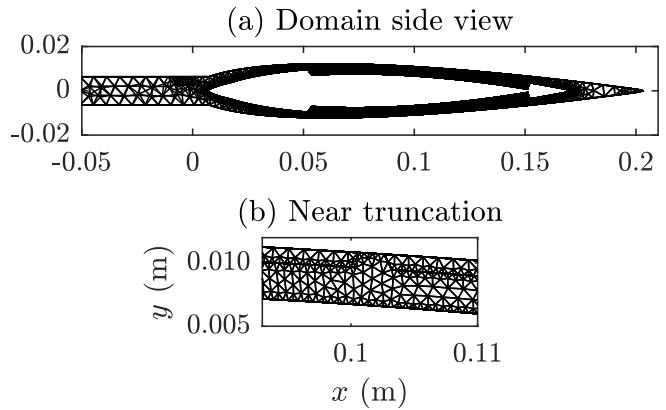}\end{center}
\caption{FEA mesh elements for (a) entire domain and (b) sub region near truncation for Case 1. All dimensions in m.}
\label{figure_comsol_mesh}
\end{figure}

The mesh resolution is a critical component affecting the accuracy of the model. Based on the sample geometry, an upper bound of $L_c$/32 = 6.35 mm is obtained by requiring at least two mesh elements to span the $h$ = $h_{\text s}$ region (Figure~\ref{figure_schematic_bc}).
A lower bound can obtained by requiring at least two elements near the truncated edge, resulting in $h_{\text t}/2$ = 0.5 mm, which also corresponds to approximately 20 elements per wavelength at the largest wavenumber of interest (Table~\ref{table_spectral}). These constraints are imposed on the mesh followed by refinement of the element growth rate and curvature until there is less than a 0.5\% change in the computed velocity over all frequencies in the domain. Figure~\ref{figure_comsol_mesh} shows a 2D slice of the mesh elements (a) over the entire domain, as well as a sub-region of the mesh (b) focusing near the truncated edge for Case 1. The tetrahedral volumetric mesh has roughly 200,000 elements with variations between cases dictated by the local geometry.

\begin{figure*}
\begin{center}\includegraphics{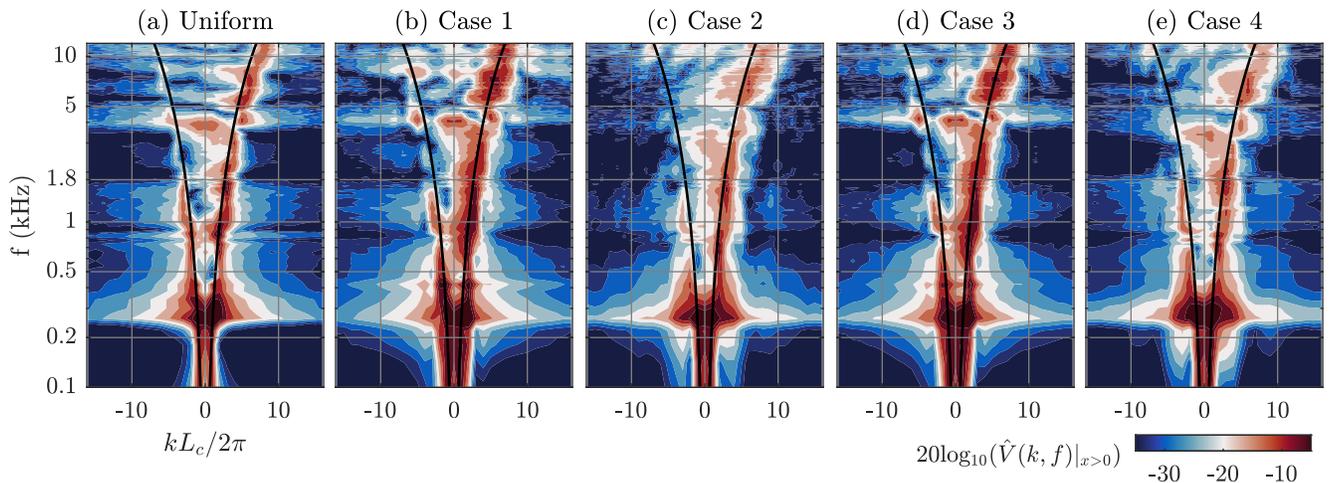}\end{center}
\caption{Contour plots of 20 log($\hat{V}(k,f)|_{x>0}$) for all the measured cases overlaid (black solid curves) with Mindlin corrected RKU composite beam model prediction for the uniformly distributed case, $k_{\text{uni}}(f)$.}
\label{figure_kw}
\end{figure*}
\section{\label{results}Results and Discussion}
\subsection{\label{kw_analysis}Wavenumber-frequency characteristics}
The time Fourier spectrum of the wavenumber-filtered velocity data is computed and as noted earlier, normalized by the corresponding value at $x/L_c$ = 0 over the entire frequency range, denoted as $\hat{V}(x,f)= V(x,f)/V(0,f)$. Distributions of the wavenumber spectrum, $\hat{V}(k,f)$ restricting to $x/L_c\geq$ 0 are shown for all cases in Figure~\ref{figure_kw}. The part of the excitation signal that goes in the other direction ($x/L_c<$ 0) couples to the mounting structure (Figure~\ref{figure_schematic_bc}). Evidently, this region has the same effect on all the samples (Figure ~\ref{figure_force}) and is thereby excluded from subsequent analysis that aims to characterize relative effects of ABH tapers on the foil. 

The Ross-Kerwin-Ungar (RKU) method is used to model the effective bending stiffness of the uniformly distributed baseline foil, assuming the TangoPlus to be a constant and thin absorbing layer on the VeroGray~\cite{lengControllingFlexuralWaves2019,climenteOmnidirectionalBroadbandInsulating2013}. The Mindlin plate correction~\cite{norrisFlexuralWavesNarrow2003} is applied to this stiffness to derive the composite wavenumber for the uniform case, denoted by $k_{\text{uni}}(f)$, also shown in Figure ~\ref{figure_kw}. Evidently, $k_{\text{uni}}(f)$ provides a very good prediction for the measured data over all the cases. Following convention, $k>0$ refers to waves traveling from the exciter (LE) towards the TE (Figure~\ref{figure_schematic_bc}), while $k<0$ contains waves reflected back from TE to LE.

The uniformly distributed case illustrated in Figure ~\ref{figure_kw}(a) has high amplitude densely concentrated around $k_{\text{uni}}(f)$ for the entire $k>0$ range over all frequencies. This is presumably because it has a constant thickness throughout the foil. Conversely, in the four ABH foils, tapers and damping layers originate and terminate at various chordwise locations, resulting in a variable local bending stiffness. Hence, as expected, the amplitude of $\hat{V}(k,f)$ is distributed over a larger region about $k_{\text{uni}}(f)$. In several regions, alternative wave-paths become available. This `smearing' of $\hat{V}(k,f)$ is also in agreement with other works~\cite{lengControllingFlexuralWaves2019,feurtadoQuietStructureDesign2017a}.

Another important conclusion from Figure ~\ref{figure_kw} is that across all the cases, reflected waves ($k<0$) symmetric to the incident waves exist only below 1.8 kHz. In other words, there is a standing wave below 1.8 kHz, corresponding to $kL_c/2\pi$ = 2.75 on the curve, or a wavelength of $L_c/17$, which is equivalent to the starting thickness, $L_c/16$ of the samples (Figure~\ref{figure_schematic_bc}). This indicates that past 1.8 kHz, waves enter all the samples, and most of them do not make it back, based on the substantially higher magnitude of the $k>0$ waves compared to those that are in the opposite direction. Therefore, subsequent discussions, where the objective remains to study the modulation of flexural characteristics in the foil regions by ABHs, are restricted to frequencies above 1.8 kHz. 

\subsection{\label{cf_analysis}Chord-frequency characteristics}
\begin{figure*}
\begin{center}\includegraphics{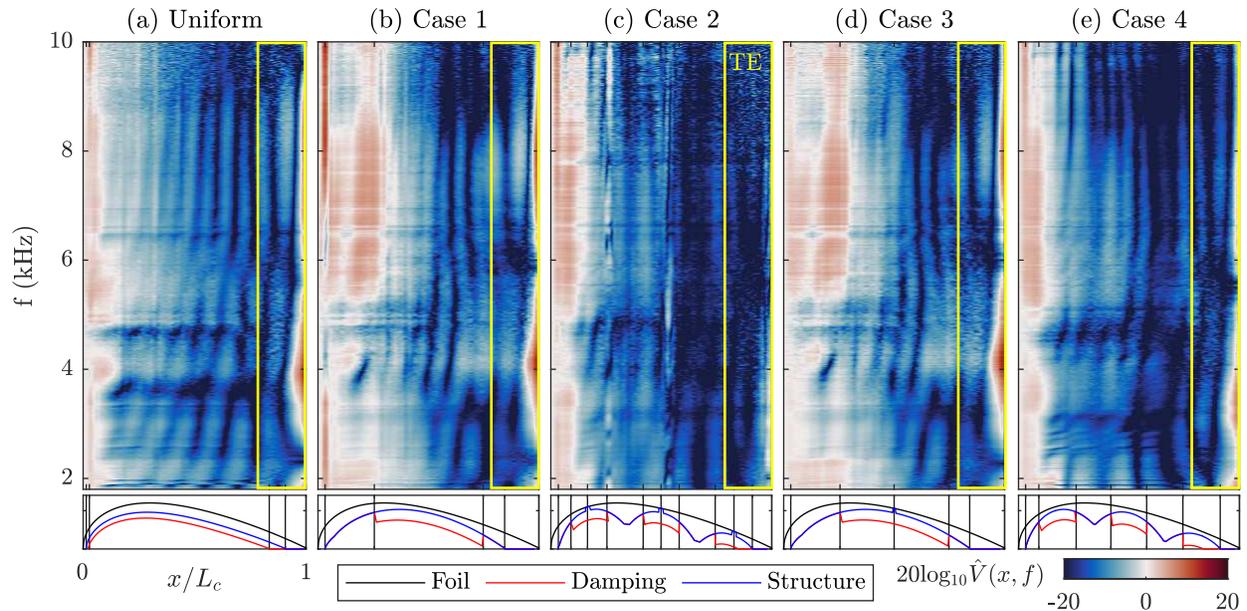}\end{center}
\caption{LDV frequency-chordwise distributions of $\hat{V}(x,f)$ in dB for all cases. Corresponding foil geometries (not to scale) below. Region 0-0.2$L_c$ close to the trailing edge (TE) across all frequencies is outlined in yellow for emphasis.}
\label{figure_fftdata}
\end{figure*}

Contour plots of  $\hat{V}(x,f)$ in dB for all the cases on top of their corresponding geometries (not to scale) are shown in Figure~\ref{figure_fftdata}. 
Chordwise locations with geometric junctions, i.e. starts and ends of tapers and damping are shown as grid lines in the bottom row to facilitate direct correlations with spatial velocity distributions. To emphasize the differences in the trailing edge vibrations between cases, the region within 0.2$L_c$ of the trailing edge is outlined in yellow. Evidently, $\hat{V}(x,f)$ varies substantially with $x/L_c$ across samples. Magnitudes are elevated near the leading and trailing edges for all cases, presumably due to the absence of damping. Lower amplitudes prevail in the mid-chord, where a series of trapped waves present as vertical bands. Some examples of this inter-junction trapping can be seen in (i) all cases above 8 kHz about their first junction, and (ii) Cases 1 and 3 in the 6-9 kHz region between the 3rd and 4th junctions. 

\begin{figure}
\begin{center}\includegraphics{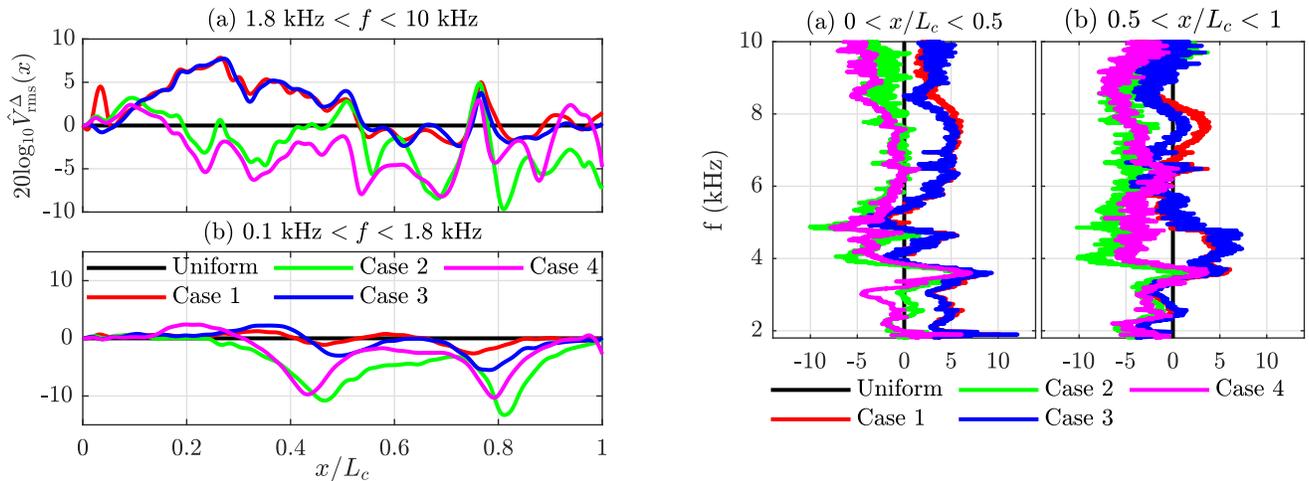}\end{center}
\caption{Profiles of $\hat{V}_{\text{rms}}^{\Delta}(x)$ in dB integrated over (a) 1.8 - 10 kHz and (b) 0.1 - 1.8 kHz for all cases.}\label{figure_fftlineX}
\end{figure}

To quantify the extent of spatial modulation across frequencies, $\hat{V}_{\text{rms}}(x)$ is computed by integrating across $f$. To facilitate comparisons, profiles of $\hat{V}_{\text{rms}}(x)$ corresponding to the baseline case are then subtracted from all the others, denoted as $\hat{V}_{\text{rms}}^{\Delta}(x)$ and shown in Figure~\ref{figure_fftlineX} expressed in dB for the (a) 1.8 - 10 kHz and (b) 0.1 - 1.8 kHz ranges.
\begin{figure}
\begin{center}\includegraphics{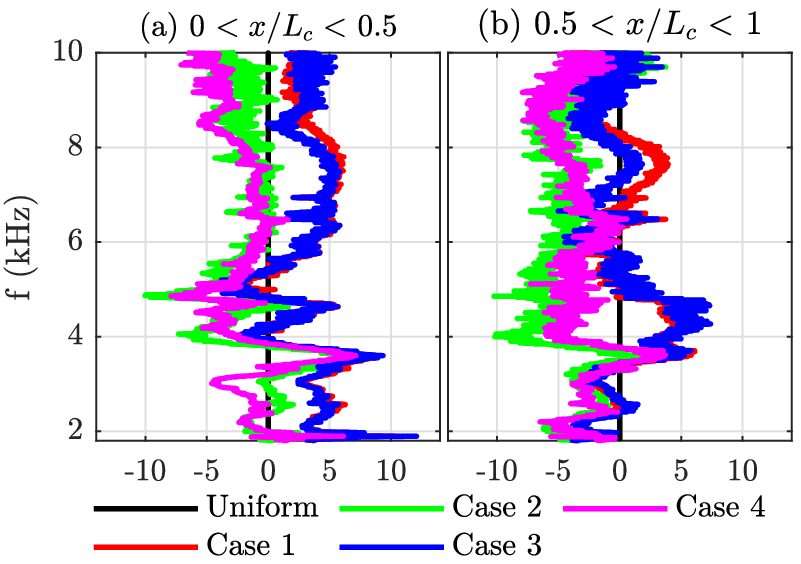}\end{center}
\caption{Baseline subtracted profiles of $\hat{V}(x,f)$ in dB averaged over the (a) first (0$<x/L_c<$0.5) and (b) second (0.5$<x/L_c<$1) halves of the chord.}
\label{figure_fftlineF}
\end{figure} 
Although frequencies below 1.8 kHz may not interact with the ABH structures (as discussed in Section~\ref{kw_analysis}), it is still interesting to note that despite everything else being similar, there is still an effect of the ABHs seen in Figure~\ref{figure_fftlineX}(b), where the $N$=3 and truncated cases have substantially lower magnitudes compared to those of the baseline, $N$=1 or continuous cases respectively. Subsequent discussion is restricted to the 1.8 - 10 kHz range.
\begin{figure*}
\begin{center}\includegraphics{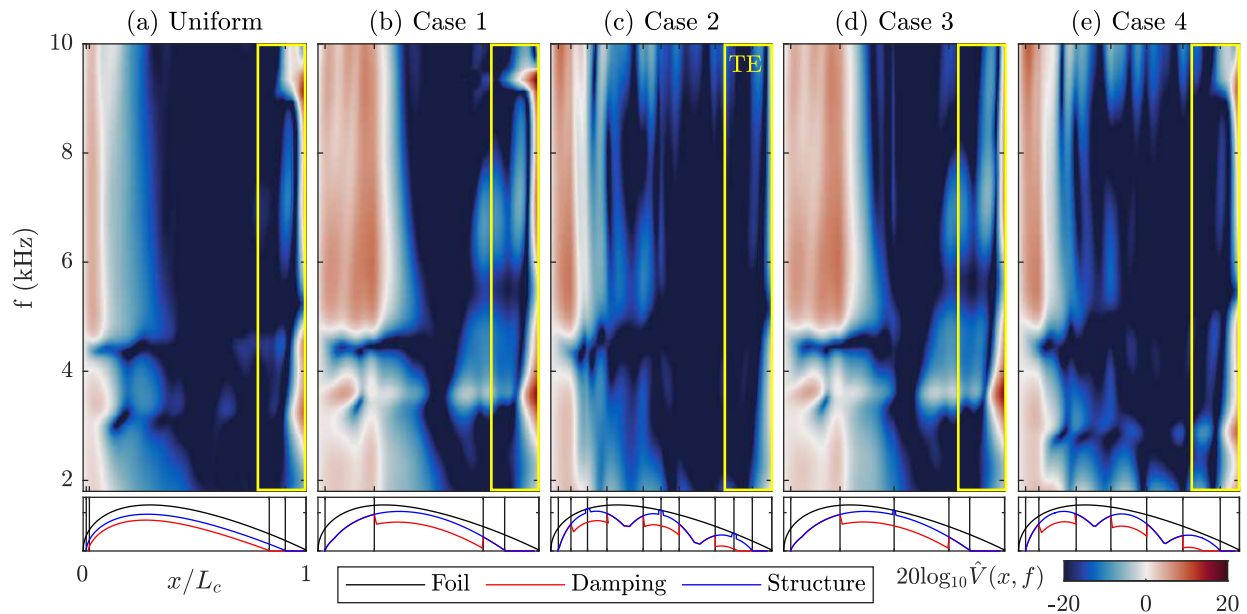}\end{center}
\caption{FEA frequency-chordwise distributions of $\hat{V}(x,f)$ in dB for all cases. Corresponding foil geometries (not to scale) below. Region 0-0.2$L_c$ close to the trailing edge (TE) across all frequencies is outlined in yellow for emphasis.}
\label{figure_comsol}
\end{figure*}

For $N$=1 geometries, i.e. Cases 1 and 3, damping only starts at $x/L_c$=0.25, thereby extending the extent of the elevated amplitude region near the leading edge. However, it is interesting to note that although damping starts at $x/L_c$=0.02 for the baseline, compared to $x/L_c$=0.08 for the $N$=3 geometries (Cases 2 and 4), elevated amplitude regions extend till $x/L_c$ = 0.16 in the 1.8-6 kHz range for all of them. Therefore, despite the absence of damping, in this frequency range, ABH tapers perform better in the 0.08 $<x/L_c<$ 0.16 region for Cases 2 and 4 when compared to the baseline. This trend reverses above 6 kHz, and also when integrated over the entire frequency range. The baseline case performs better than all the current cases (Figure~\ref{figure_fftlineX}) near the leading edge, where the earlier onset of damping outweighs all other factors. 

In the mid-chord region (0.2 $<x/L_c<$ 0.8), there is a clear distinction between different cases. $N$=3 geometries substantially outperform the baseline case throughout, in fact the reduction in amplitude is above 6-7 dB at $x/L_c$ = 0.25, 0.35, 0.65 and 0.7 as seen in Figure~\ref{figure_fftdata} and~\ref{figure_fftlineX}). $N$=1 geometries perform consistently worse than the baseline by as much as 5-7 dB at $x/L_c$ = 0.2-0.35. The only location where $N$=1 geometries fair better than the baseline, by 2-3 dB, is around $x/L_c$=0.6 and 0.7.

Near the trailing edge (0.95 $<x/L_c<$ 1), $N$=1 geometries perform very similar to the baseline.  $\hat{V}_{\text{rms}}(x)$ as well as the spatial distributions along the entire frequency range (Figure~\ref{figure_fftdata}) are very similar. However, $N$=3 geometries showcase substantial reduction in magnitude, as high as 10 dB for the truncated Case 2, when integrated over the entire frequency range (Figure~\ref{figure_fftlineX}). This result can have significant implications for airfoil design where trail-edge noise control is of interest. The spatial velocity distributions (Figure~\ref{figure_fftdata}) reveal that there is not only a broadband reduction in amplitude near the trailing edge, but also a down-shift in the 4 kHz baseline peak to around 3.3 kHz. 

To further quantify the frequency modulation by the ABHs, Figure~\ref{figure_fftlineF} shows the baseline subtracted $\hat{V}(x,f)$ in dB averaged in the (a) first (0$<x/L_c<$0.5) and (b) second (0.5$<x/L_c<$1) halves of the chord. As discussed previously, there is a front-loading effect prevalent in the $N$=1 cases, that is also highlighted by Figure~\ref{figure_fftlineF}(a), with a 5 dB increase in velocity above 6 kHz. However, for the second half of the foil (Figure~\ref{figure_fftlineF}(b)) in this same range, Cases 1 and 3 exhibit a 3 dB reduction compared to the baseline. Therefore, the same frequency range is modulated 5 dB above and 3 dB below the baseline in the first and second halves respectively by the $N$=1 geometries. The $N$=3 cases 2 and 4 show 3-5 dB reduction in the first half of the foil for some frequency bands, i.e. 4-5 and 8-10 kHz while performing similar to the baseline in other regions of the leading edge half. However, in the second half (Figure~\ref{figure_fftlineF}), they exhibit a 5 dB reduction in amplitude across 4-10 kHz range. Such a large broadband reduction in amplitude across the entire second half of the airfoil can have significant implications to applications involving control of flow separation, stall and other transitional and unsteady effects.
\subsection{\label{fea_results}FEA results}
Distributions of $\hat{V}(x,f)$ in dB obtained from the FEA simulations for all the cases along with their corresponding geometries are shown in Figure~\ref{figure_comsol}. The plot extents, color map range, grid and placement are identical to those of Figure~\ref{figure_fftdata} to facilitate a comparison of the simulations with the LDV measurements. There is excellent agreement between the FEA model and the LDV measurements. They capture the spatial extents of the elevated amplitude regions near the leading and trailing edges, as well as the differences between cases across the entire frequency range. For e.g. the 4 kHz peak and neighboring features near the leading edge for the baseline and Cases 1 and 3, albeit shifted slightly, match very well in the FEA results. The trailing edge distributions here also show the close similarities between the baseline and $N$=1 geometries, whereas the $N$=3 cases also show a significant broadband reduction in amplitude, in good agreement with the LDV data. Furthermore, the inter-junction trapping examples (i)-(ii) highlighted for the LDV data are also captured extremely well by the FEA models. The frequency down-shift of the trailing edge 4 kHz peak from the baseline to the $N$=3 cases is also evident.

There are some minor discrepancies that are worth mentioning. First, as eluded to earlier, the entire data, across all samples appears slightly downshifted in frequency when compared to the measurements. This suggests that the printed samples had slightly different material properties than those used in the FEA model. This difference is small enough that there doesn't seem to be any significant change in vibrational modes. Second, not all the features in the mid-chord, especially above 5 kHz for the baseline case are captured by the FEA. Third, Cases 2 and 4 appear to have elevated amplitudes in the FEA results (Figure~\ref{figure_comsol}(c) and (e)) above 8 kHz. This might be linked to the frequency down-shift described earlier, which carries through the entire range. Since this is a wavenumber-effect, the shift is expected to increase with increasing frequency. Thereby, the elevated amplitudes around 9 kHz would be expected around 10-11 kHz in the LDV data. As evident from Figure~\ref{figure_force}, the measured value drops significantly in this range until at least 12 kHz, where the measured LDV signal has a low signal-to-noise ratio, affecting the frequency normalization and precluding any conclusions to be drawn in this range. Despite these issues, the FEA model does a really good job at capturing the key features of the velocity distributions and most importantly, the differences between geometries. Therefore, for a given desired objective function, this can help reduce the fabrication requirement substantially.

\begin{figure}
\begin{center}\includegraphics{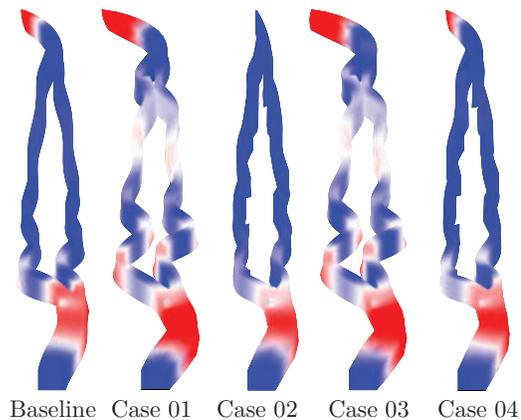}\end{center}
\caption{FEA displacements at 4 kHz for all samples amplified by 30,000 times.}
\label{figure_comsol_snapshot}
\end{figure}

Furthermore, the FEA model can be used to visualize the ABH effect by looking at frequency-specific displacement fields for all cases. Figure~\ref{figure_comsol_snapshot} shows a snapshot of the displacement, amplified by 30,000 times, at 4 kHz for all the samples obtained from the FEA models.
The leading and trailing edges of the $N$=1 cases deflect even higher than the baseline. However, $N$=3 geometries, as seen in all the data, have substantially lower amplitudes throughout the foils, especially at the trailing edge, elucidating the overall impact of the ABHs.

\subsection{\label{parametric_study}ABH-foil parametric study}
Although a comprehensive optimization for various cost functions and parametric spreads remains out of the present scope, a glimpse of the spatial vibrational modulation that can be achieved by implementing ABHs in airfoils is provided here. The validation of the FEA models also confirms the material properties moving forward. As discussed earlier, FEA models were required to be 3D to match better to the fabricated samples. However, given that most applications of foils are expected to have a large span-to-chord ratio, a 2D plate approximation may be sufficient. This also makes FEA computation for the entire shortlisted LUT feasible, which is automated using COMSOL Livelink with MATLAB.

\begin{figure}
\begin{center}\includegraphics[width=\linewidth]{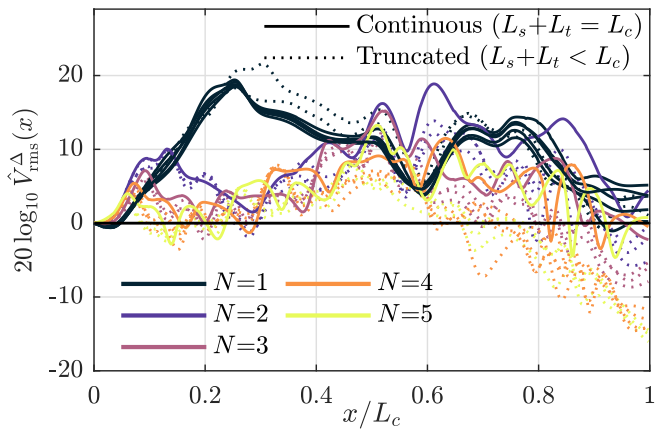}\end{center}
\caption{ABH foil parametric study: 2D FEA-generated profiles of $\hat{V}_{\text{rms}}^{\Delta}(x)$ in dB for shortlisted cases.}
\label{figure_parametric_1D}
\end{figure}

For the present purpose, 24 geometries with masses within 2\% of each other are shortlisted from the LUT (Table~\ref{table_pars}). For brevity, we restrict only to whole-numbered $N$ values, i.e. the geometries terminate with VeroGray. As noted earlier, the LUT and shortlist thresholds can be made arbitrarily dense to accommodate the objective at hand. Figure~\ref{figure_parametric_1D} contains profiles of $\hat{V}_{\text{rms}}^{\Delta}(x)$ in dB for all 24 geometries.
They are colored based on their $N$ values, while continuous and truncated geometries are represented as solid and dotted lines respectively. Curves also represent variations in $n$, $L_{\text s}$, $L_{\text t}$, $L_{\text d}$, $h_{\text s}$ and $h_{\text d}$ while satisfying the mass constraints, and these have not been explicitly identified to restrict to the present scope. Evidently, the profiles vary as high as 20 dB compared to the baseline case. As noted in the fabricated cases earlier, even here, it appears that increasing $N$ reduces the overall vibration level, especially in the second half of the chord, closer to the trailing edge. This is presumably due to a compounding effect as the waves encounter more damped ABHs before reaching the trailing edge. Conversely, $N$=1 cases, have the most elevated vibration levels, especially in the first half of the chord. Truncation also reduces the amplitude, again, more so, closer to the trailing edge. 

\section{\label{conclusions} Conclusions}

This study introduces a framework for the design and implementation of ABHs in airfoils. A multi-parameter damped-ABH generation function is mapped onto a NACA series airfoil. Four ABH geometries and a uniformly distributed baseline, all with the same mass of structure and damping are fabricated using multi-material PolyJet 3D printing. Laser Doppler vibrometer measurements of velocity along the airfoil chord in response to a broadband 0.1 - 12 kHz excitation are performed for all the cases. 3D FEA is also performed on the fabricated geometries, to enable for model and material property validation. Furthermore, a parametric 2D FEA study is performed on shortlisted geometries using the validated material properties, to showcase the mitigation and modulation that is achievable by implementing ABH design. Key findings of the study are described below, 
\begin{itemize}
\item Wavenumber-frequency characteristics of the measured data follow the Mindlin-corrected RKU model. The uniform baseline is densely concentrated about the curves, whereas all the ABH cases show $k$-space smearing of energy in agreement with findings from other ABH studies~\cite{feurtadoQuietStructureDesign2017a}. 
\item In general, spatial distributions of velocity as a function of frequency, normalized by that at excitation reveal substantial variations between samples. Magnitudes are elevated near the leading and trailing edges of the foils, while lower amplitudes prevail in the middle. In the ABH cases, a series of standing waves are trapped between local junctions where tapers and damping transition.
\item In comparison to the uniform baseline, there is an reduction of 5 dB in the magnitude across the entire frequency range for foils with $N$=3 embedded ABHs on average over the chord length, with up to a 10 dB reduction near the trailing edge for the truncated case. On the other hand, ABH foils with $N$=1 ABHs are associated with an increase in magnitude by as much as 5-7 dB in the first half of the chord, while remaining comparable to the baseline towards in the second half.
\item Baseline-subtracted velocity profiles averaged in the first half of the chord show that the front-loading effect for $N$=1 ABH cases exists above 6 kHz. Profiles in the second half elucidate a broadband (4-10 kHz) reduction in amplitude by 5 dB with the $N$=3 cases. 
\item 3D finite element models of the five samples are in good agreement with LDV measurements. They capture the spatial extents of the leading and trailing edges as well as inter-sample variations very well. These validate the model as well as the material properties.
\item Two-dimensional parametric FEA results indicate that a modulation in the velocity amplitude of as much as 20 dB with ABH-embedded foil designs. The effects of the number of ABHs and truncation are in agreement with the trends measured in the experiments.
\end{itemize}
In conclusion, this study provides an insight into the design, fabrication, performance, modeling and optimization of ABH-embedded airfoils. Given a constant mass structure, airfoils can be designed to mitigate, focus or modulate vibrations for any chordwise region by adapting the process presented in this study.  In applications where minimizing the noise radiation from trailing edges is of concern, findings from the present study can be used to achieve upwards of a 10 dB reduction in vibration. For cases where minimizing broadband vibrations for structural integrity and wear of foils is important, multiple truncated ABHs can be distributed in a spanwise orientation. Alternatively, for achieving flow control at specific chordwise locations or frequency bands, using the present framework, energy can be added or subtracted as desired from the boundary layer close to the foil leading to superior performance and efficiency. Furthermore, other applications involving energy harvesting or restructuring can benefit from the front-loading effect introduced by the $N$=1 ABH cases.
\begin{acknowledgments}
This work was supported by the Office of Naval Research.
\end{acknowledgments}
\clearpage
\bibliographystyle{unsrt}	
\bibliography{References}
\end{document}